\documentclass[prb,aps,twocolumn,showpacs,showkeys]{revtex4}

\usepackage{amssymb}
\usepackage{graphicx}

\begin{document}

\title[Self-assembled aggregates in the gravitational field]{Self-assembled
aggregates in the gravitational field:\\
growth and nematic order}
\author{Vladimir A. Baulin\footnote{\normalsize{e-mail:
vbaulin@cea.fr\\ \\ J. Chem. Phys., 119(5), 2874-2885 (2003).}}}

\affiliation{Service des Interfaces et des Mat\'{e}riaux
Mol\'{e}culaires et Macromol\'{e}culaires, DRFMC, CEA-Grenoble, 17
rue des Martyrs, 38054 Grenoble, Cedex 9, France}



\begin{abstract}
The influence of the gravitational field on the reversible process
of assembly and disassembly of linear aggregates is focus of this
paper. Even the earth gravitational field can affect the
equilibrium properties of heavy biological aggregates such as
microtubules or actin filaments. The gravity gives rise to the
concentration gradient which results in the distribution of
aggregates of different lengths with height. Strong enough
gravitational field induces the \textit{overall growth} of the
aggregates. The gravitational field facilitates the isotropic to
nematic phase transition reflecting in a broader transition
region. Coexisting phases have notedly different length
distributions and the phase transition represent the interplay
between the growth in the isotropic phase and the precipitation
into nematic phase. The fields in an ultracentrifuge can only
reinforce the effect of gravity, so the present description can be
applied to a wider range of systems.
\end{abstract}

\maketitle

\section{Introduction}

Living polymerization has a substantial number of industrial
applications in the production of soaps, dyes, organometallic
complexes, \textit{etc.}\cite{catescando,Greer} Moreover, some
biological objects owe their functionality to reversible processes
of assembly and disassembly. Particular examples are microtubules,
actin and intermediate filaments composing the cytoskeleton of
living cells.\cite{Shadows,DogteromLes,Dogterom} They are involved
in several vital processes such as transport, chromosomal
segregation,\cite{mitosis} growth and division of a cell. In
contrast to common synthetic living polymers, biological
self-assembled objects have a considerable molecular masses,
sufficient to induce a significant concentration gradient in an
aqueous suspension of these objects. In turn, the concentration
gradient can influence the equilibrium size of the objects and for
long rigid aggregates it can even provoke the isotropic -- nematic
phase transition.

Before we address such phenomena it is useful to get an estimate
of the relevance of the effect of gravity for a real system. Let
us focus on a particular example of microtubules. The typical
length of a microtubule is $10$ $\mu $m,\cite{Panda,Derry} the
linear density is $1.6\times 10^{5}$ (g/mol)/nm, so the total mass
of a microtubule is $\sim 10^{9}$ g/mol. The molecular mass of a
tubulin dimer, subunit of a microtubule, is $m_{0}\sim 10^{5}$
g/mol,\cite{Shadows} thus, one microtubule comprises of $N\sim
10^{4} $ tubulins. The influence of the gravity is significant if
the potential energy of the object in the gravitational field is
of order or higher the thermal energy, $kT$, \textit{i.e.} if
$(mgh/kT)N\equiv \beta N\gtrsim 1$, where $m=m_{0}-\widetilde{\rho
}\upsilon $ is the mass of a subunit corrected for buoyancy
($\widetilde{\rho }\sim 1$ g/cm$^{3}$ is the density of water,
$\upsilon \sim 100$ nm$^{3}$ is the subunit volume, so the
effective mass in a solution is $m\sim 10^{4}$ g/mol). Thus, the
gravity is important when $\beta \gtrsim \beta ^{\ast }=1/N$.
Assuming the total height of the vessel, $h\sim 1$ cm, the
gravitational acceleration, $g\sim 10$ m/s$^{2}$, the temperature,
$T\sim 300$ K, we get $\beta \sim 10\beta ^{\ast }$. So, even the
earth gravitational field can influence the equilibrium properties
of such heavy objects as microtubules. Note, that one can reach
$10^{4}-10^{5}$ times larger gravitational acceleration in an
ultracentrifuge, \cite{Anacker} which sometimes is used in the
process of preparation of microtubules,\cite{Derry} or in the
pelleting experiments. \cite{Shah,Hlavanda} To this end, the
gravitational field can serve as a driving force for the assembly
and, eventually, to the nematic ordering of growing aggregates.
Another way to study the equilibrium distribution of linear
aggregates is the rotating clinostat experiments\cite{clinostat}
which constitute the rotation of a tube about the horizontal axis.
Changing the direction of the "fall" all the time can effectively
reduce the gravity. Such apparatus gives the gravitational fields
in the range $10^{-2}\div 10^{2}$ g.

The rotating clinostat experiments\cite{clinostat} were used to
examine the microtubule self-organization and structuration by
reaction-diffusion mechanisms leading to different pattern
formation. It was shown, that microtubules self-organize in
different patterns in the normal gravitational field, while no
self-organization occurs in the absence of the
gravity.\cite{Tabony,tabony2,tabony3}

It is noteworthy, that the gravity can also bring about the
opposite process, namely the floating to the surface when
$m_{0}<\widetilde{\rho} \upsilon$. One of the most known processes
of that type is the creaming of emulsions. In our consideration
this only changes the sign of the effective mass, $m$, while the
top and the bottom switch their places.

The problem of the influence of weak gravitational fields on
different vital processes is largely investigated in the framework
of so-called microgravity research performed in sounding rockets,
parabolic flights or a space station. In particular, it concerns a
number of experiments conducted on board of space-shuttle flights.
For example, the formation of liposomes is altered by the
gravitational field:\cite{liposomes} the size of the liposomes
formed in the absence of the gravity is considerably larger than
that formed in the normal earth conditions. The gravity also
influence the long-range ordering of some systems:\cite{micrograv}
space-grown crystals have better optical quality, smoothness of
crystal faces, \textit{etc}. The problem of crystal growth under
gravity is also discussed in ref. \onlinecite{crystal}.

However, the number of theoretical treatment of these observations
is not so impressive. Here are some examples: the size
distribution with height of noninteracting linear micelles in the
gravitational field has been analyzed in ref.
\onlinecite{Duyndam}. It was shown, that the concentration
gradient due to gravitational or centrifugal field leads to the
hyperbolic dependence of the average aggregation number with
height in contrast to simple barometric dependence for
nonaggregating objects. The dynamics of linear micelles in the
ultracentrifuge is studied in ref. \onlinecite{Duyndam2}. The
sedimentation of colloidal rods in an ultracentrifuge, has been
studied theoretically in ref. \onlinecite{Fraden}.

In our previous paper\cite{me} we studied the isotropic to nematic
phase transition of rigid monodisperse rods with \textit{fixed
size }in the gravitational field. The major result is that the
gravity can induce the nematic ordering in the system of rigid
rods, while the main effect of the gravity on the phase diagram is
the broadening of the phase coexistence region. It was also shown
that interactions between rigid rods can lead to deviations from
the barometric distribution of rods with height.

In the present paper we analyze a wider range of systems. Instead
of fixed length of rods,\cite{me} we assume that the size of rods
determines by the equilibrium conditions. In particular, this is
the case of (i) living polymers, which can attach monomers to the
end of the chain, (ii) linear surfactant micelles, (iii)
biological self-assembled objects such as microtubules and actin
filaments. Although all these systems have different geometries
and different size scales, they have additional degree of freedom,
namely the annealed length of the aggregates which can be tuned by
external conditions. Thus, we expect, that the gravitational field
can have a severe influence on the equilibrium size and the
concentration profile of the aggregates. The new result of this
paper is that the gravity induces the overall growth of the
aggregates.

Concerning the isotropic to nematic phase transition of stiff
aggregates with annealed size, we expect the same effect of the
gravity as in the case of rods with fixed length. Namely, the
gravity can induce the nematic ordering and broaden the
coexistence region. However, annealed length implies a large
polydispersity of the aggregates, thus, we investigate the
influence of the gravity on the distribution of lengths in a
solution. Since the phase separation is coupled with the
redistribution of aggregates of different lengths with different
heights, we discuss the structure of the two coexisting phases and
the distribution of the average aggregation number with height.

The paper is organized as follows. In the next section the general
definitions and the expressions for the free energy and the
distribution function are given. In the section III we consider
the equilibrium properties of the homogeneous isotropic solution
of living polymers under gravity. The concentration profiles and
the average aggregation numbers as a functions of height are
analyzed. In the section IV, we examine the isotropic to nematic
phase transition in the gravitational field. We discuss the effect
of the broadness of the phase separation region in the
gravitational field for living polymers, the length distribution
functions of the aggregates in the coexisting region and
polydispersity of the phases. In the Conclusion we summarize the
obtained results.

\section{Free energy and distribution function}

Living polymers are linear aggregates which can bind reversibly
one subunit to the end of the chain. They show a broad
distribution of lengths, since their energy depends only on end
effects and there are no packing constrains preventing the
infinite growth as in the case of spherical micelles. The scission
end energy (in units of $kT$), $\delta >0$ ensures energetically
favorable formation of aggregates. The aggregation energy per subunit, $%
\varepsilon _{N}$ has a simple form:\cite{Safran} $\varepsilon
_{N}=\varepsilon _{\infty }+\delta /N$, where $\varepsilon
_{\infty }$ is the bulk energy, which is independent on $N$. Thus,
$N\varepsilon _{N}$ is the energy of an aggregate with the
aggregation number $N$. The subunit concentration is assumed to be
much larger than \textit{CMC}, so the aggregates are rather long.

Our description is based on the theories of living polymers and
the liquid
crystalline transition in the absence of the field.\cite%
{OdijkRev,taylor,gelbart1,
gelbart2,McMullen,Cuesta,schootmain,schootgrowth,
schootphase,schootisotropic,schootnem} For our description we take
as a starting point ref. \onlinecite{schootmain}. It is known,
that the assumption of perfectly rigid aggregates leads to
unphysically strong coupling of growth and nematic
ordering.\cite{Odijk,schootmain,schootphase}

That is why, we have to allow for the partial flexibility of the
aggregates. We do it in accordance to
refs.\cite{Odijk,schootisotropic,schootphase}, where the
flexibility of the chains is coming through a particular form of
the trial function in the nematic phase. However, in this approach
the flexibility of the isotropic phase is not taken into account.

We consider the solution of living polymers in the vessel subject
to the gravitational field. Suppose, that the gravitational field
acts in the $z$ direction, such that the subunit concentration
varies from its maximal value at $z=0$ (bottom) to its minimal
value at $z=1$ (top). Our discussion is focused on the case of
long aggregates. We ignore the end effects, though they can be
important for short chains\cite{schootmain} that dominate at the
top of the vessel, but for the subunit concentration well above
the \textit{CMC}, the fraction of short aggregates in a solution
is low.

In the following we will see, that the gravity can lead to the
nematic ordering in the described system, thus, we write the free
energy in a general form which assumes the orientational order.
The free energy can be written in terms of the dimensionless
concentration of rods each comprised of $N$ subunits oriented in
the spatial angle $\Omega $ on the height $z$ as $C(N,\Omega ,z)$
(this is the number concentration multiplied by the volume of a
single subunit, $\upsilon $). $C(N,\Omega ,z)$ obeys the following
constraint

\begin{equation}
\int NC(N,\Omega ,z)dNd\Omega =\phi (z)  \label{constr}
\end{equation}
where $\phi (z)$ is the total volume fraction of subunits on the
height $z$. The total subunit concentration is $\Phi =\int \phi
(z)dz$.

The free energy is composed of an ideal and an interaction terms, $%
F=F_{id}+F_{int}$. The ideal term per unit volume is written in
the form

\begin{eqnarray}
\frac{F_{id}(z)}{kT}&=&\int C(N,\Omega ,z)\left[ \ln \left( 4\pi
C(N,\Omega ,z)\right) -1 +\right. \nonumber \\
&&\left. N\varepsilon _{N}\right] dNd\Omega +\int C(N,\Omega
,z)\beta NzdNd\Omega  \label{Fid}
\end{eqnarray}

The first term is the well known expression for the free energy of
self-assembled aggregates.\cite{Safran} The factor $4\pi $ is
introduced for convenience of description of the isotropic phase,
though it can be absorbed in the energy of an aggregate,
$N\varepsilon _{N}$. The last term is the potential energy in
units $kT$ of aggregates in the gravitational field. The only
parameter associated with the field is $\beta =(mgh)/kT$, which
depends on the total height of the vessel. Note, that although the
last term in (\ref {Fid}) coincide with that in the case of
monodisperse rods of fixed length,\cite{me} the definitions of
this parameter does not coincide. Here $m$ is the mass of a single
subunit, while in ref. \onlinecite{me} $m$ is the total mass of a
rod. Again, we assume that the characteristic length of the
variation of the field is much bigger than the length of a rod,
although the fields in the ultracentrifuge can be very high and
this assumption can be violated.

In order to apply the Onsager approximation,\cite{Onsager,Odijk}
we assume excluded volume interactions only. Moreover, only pair
interactions are taken into account, higher order interaction
terms are neglected. This is a strong approximation which is
strictly speaking not valid for high concentrations in the nematic
phase. As we will see, the strong gravitational field induces the
precipitation of extremely long, and thus, heavy, aggregates and
the second virial approximation should be corrected.

Nevertheless, this approximation works well in the isotropic phase
and it also gives reasonable qualitative estimates in the nematic
phase.

The interaction term per unit volume for athermal aggregates,
written in the
second virial approximation has the following form\cite%
{schootmain,schootisotropic}

\begin{equation}
\frac{F_{int}(z)}{kT}=\frac{1}{\upsilon }\int C(N,\Omega
,z)C(N^{\prime },\Omega ^{\prime },z)B_{2}(\gamma )dNdN^{\prime
}d\Omega d\Omega ^{\prime } \label{Fint}
\end{equation}

where $B_{2}(\gamma )$ is the second virial coefficient which
depends on the angle $\gamma $ between two rods. Neglecting the
end effects, it can be written in the form\cite{Odijk,Onsager}
$B_{2}(\gamma )=LL^{\prime }D\left\vert \sin \gamma \right\vert $,
where $L$ and $L^{\prime }$ are the lengths of two cylinders, $D$
is their diameter. We can rewrite this expression in terms of
aggregation numbers $N$, $N^{\prime }$ and the volume of a
subunit, $\upsilon $, as $B_{2}(\gamma )=\varkappa \upsilon
NN^{\prime }\left\vert \sin \gamma \right\vert $, where $\varkappa
$ is the parameter associated with the \textit{form} of a subunit.
In the case of cubic subunits, $\varkappa =1$, while for
cylindrical subunits, $\varkappa =4l/D$, where $l$ is the height
of a cylinder. However, to apply this notation to the case of
microtubules, we should note, that a microtubule has complex
geometry. Microtubules are hollow cylinders consisting of $f=13$
columns of tubulin on the surface.\cite{Shadows} So, they grow up
at $f$ places at the same time. Thus, the length of a microtubule
$L\sim fN$ and $\varkappa =4lf^{2}/D$. In the following we assume
$\varkappa =1$ for simplicity.

Variation of the free energy (\ref{Fid}), (\ref{Fint}) with
respect to the
concentration $C(N,\Omega,z)$ along with the constraint condition (\ref%
{constr}) leads to the following expression

\begin{eqnarray}
C(N,\Omega ,z)&=&\frac{1}{4\pi }\exp \left( N(\mu -\varepsilon
_{N\ }-\beta z)- \right. \nonumber \\
&& \left. \frac{2}{\upsilon }\int C(N,\Omega ,z)B_{2}(\gamma
)dNd\Omega \right) \label{CNOm}
\end{eqnarray}
where $\mu $ is the Lagrange multiplier, which can be eliminated
with the help of the constraint (\ref{constr}).

Introducing the average aggregation number of rods in the
direction $\Omega$ at height $z$ as

\begin{equation}
\overline{N}(\Omega,z)=\frac{\int NC(N,\Omega ,z)dN}{\int
C(N,\Omega ,z)dN} \label{N}
\end{equation}
we can write the expression for the distribution function
$C(N,\Omega ,z)$ in a simpler form

\begin{eqnarray}
C(N,\Omega,z) &=&\frac{1}{4\pi }\exp \left(
-\frac{N}{\overline{N}(\Omega,z)}
-\delta \right)  \nonumber \\
&=&\frac{\phi }{\int \overline{N}^{2}(\Omega,z)d\Omega }\exp
\left( -\frac{ N }{\overline{N}(\Omega,z )}\right)  \label{C}
\end{eqnarray}
To write the last equality we used the constraint condition
(\ref{constr}).

The final expressions could be considerably simplified if we note,
that (i) there is no angular dependence in the isotropic phase,
(ii) in the nematic phase we can split the distribution function
$C(N,\Omega ,z)$ in two parts,\cite{McMullen} one dependent on the
aggregation number, $N$, and another dependent on spatial angles,
$\Omega $. Namely, we denote the length distribution function,
$P(N,z)=\int NC(N,\Omega ,z)d\Omega $ with the normalization
condition $\int P(N,z)dN=\phi (z)$ and the orientation
distribution function $R(N,\Omega ,z)=\frac{NC(N,\Omega ,z)}{\int
NC(N,\Omega ,z)d\Omega }$ with the normalization $\int R(N,\Omega
,z)d\Omega =1$. With this, $C(N,\Omega
,z)=\frac{P(N,z)}{N}R(N,\Omega ,z)$ and the free energy,
$F=F_{id}+F_{int}$, now yields in the form

\begin{eqnarray}
&&\frac{F(z)}{kT} =\int \frac{P(N,z)}{N}dN\left[ \sigma (N,z)+\ln
\frac{P(N,z)}{N}-1+\right.  \nonumber \\
&&\left. N\varepsilon _{N}+\beta Nz+\frac{\pi }{4}N\int
P(N^{\prime },z)\rho (N,N^{\prime },z)dN^{\prime }\right]
\label{F}
\end{eqnarray}
where $\sigma (N,z)=\int R(N,\Omega ,z)\ln 4\pi R(N,\Omega
,z)d\Omega $ is the orientational entropy, $\rho (N,N^{\prime
},z)=\frac{4}{\pi }\int R(N,\Omega ,z)\\R(N^{\prime },\Omega
^{\prime },z)\left\vert \sin \gamma \right\vert d\Omega d\Omega
^{\prime }$ is the excluded volume term. We keep this nomenclature
in order to be consistent with refs.\cite{me,schootmain,
schootphase,schootisotropic,Onsager,Odijk} In the isotropic phase
$\sigma _{i}=0$, $\rho _{i}=1$ and $R_{i}(N,\Omega ,z)=1/4\pi $,
so the expression for the free energy simplifies significantly.

In the next section we describe the concentration profile and the
average length of the aggregates assuming that the solution of
rods is homogeneous and isotropic for the whole range of $z$.

\section{Homogeneous solution: growth of aggregates}

Since the directions of rods in the isotropic phase are completely
disordered, the distribution function does not depend on angles, $
C_{i}(N,\Omega )=C_{i}(N)$ and eqs. (\ref{CNOm}), (\ref{N})
give\cite{foot1} the well known result for the average aggregation
number in the isotropic phase $\overline{N}_{i}(z)=\sqrt{\phi
_{i}(z)e^{\delta }}$. In turn, the free energy expressed in terms
of the length distribution function, $P(N,z)$ (\ref{F}) leads to
the free energy of the isotropic phase in the form

\begin{equation}
\frac{F_{i}(z)}{kT}=\phi _{i}(z)\left[ \varepsilon _{\infty }+\beta z-\frac{2%
}{\overline{N}_{i}(z)}+\frac{\pi }{4}\phi _{i}(z)\right]
\label{fi}
\end{equation}
Here we used again the phenomenological expression for the
aggregation energy (per $kT$) of linear aggregates,\cite{Safran}
$\varepsilon _{N}=\varepsilon _{\infty }+\delta /N$. The chemical
potential $\mu _{i}(z)=\partial F_{i}(z)/\partial \phi _{i}(z)$ is

\begin{equation}
\frac{\mu _{i}(z)}{kT}=\varepsilon _{\infty }+\beta
z-\frac{1}{\overline{N} _{i}(z)}+\frac{\pi }{2}\phi _{i}(z)
\label{mui}
\end{equation}
and the osmotic pressure $p_{i}(z)=1/\upsilon \phi
_{i}^{2}\partial \left[ F_{i}/\phi _{i}\right] /\partial \phi
_{i}(z)$ is

\begin{equation}
\frac{p_{i}(z)}{kT}\upsilon =\phi _{i}(z)\left[
\frac{1}{\overline{N}_{i}(z)} +\frac{\pi }{4}\phi _{i}(z)\right]
\label{pi}
\end{equation}

Note, that these expressions formally have the same form as in the
absence of gravity\cite{schootmain,schootisotropic} (the only
difference is that the concentration $\phi _{i}$ depends on $z$).
This is because on the lengths smaller than the variation of
gravitational field, the system does not "feel" the field, since
all aggregates have almost the same potential energy. In our
approximation the gravitational field influence the system
only through the concentration gradient. The expressions (\ref{mui}), (\ref%
{pi}) will be of use to the phase coexistence.

In the equilibrium the chemical potential does not depend on height. Thus, $%
\mu _{i}(z)=const$, eq. (\ref{mui}), determines the concentration
profile induced by the field. In turn, the variation of the
concentration with height leads to a redistribution of aggregates
with different lengths at different heights. To address this
issue, we focus on two limits: ideal aggregates (low
concentrations) and considerably long aggregates (high
concentrations).

\begin{figure}
\begin{center}
\includegraphics[height=5.7cm,width=8cm]{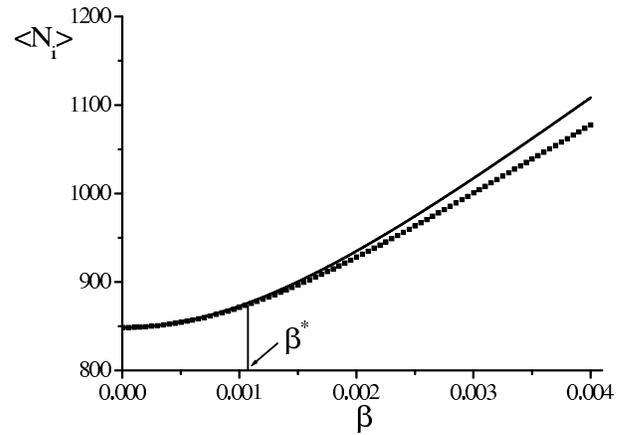}
\end{center}
\caption{Average aggregation number on the whole range of heights,
$\langle N_{i}\rangle $, of ideal linear aggregates (\ref{Niid})
as a function of a dimensionless parameter associated with the
gravitational field, $\beta =mgh/kT$ (solid line) in comparison
with the exact solution for nonideal aggregates (squares). The
parameters used: $\delta =25$, $\Phi =10^{-5}$ (low concentration
limit). The crossover between the week field and the strong field,
which induce considerable growth of aggregates is $ \beta ^{\ast
}\sim 1/\overline{N}_{i,0}$.}
\end{figure}

In the limit of noninteracting \textit{ideal} aggregates in the
gravitational field,\cite{Duyndam} the last term in the expression
for chemical potential (\ref{mui}) arising from the excluded
volume interactions is eliminated. This leads to the following
concentration profile

\begin{equation}
\phi _{i}(z)=\frac{\phi _{i,0}}{\left( 1+\sqrt{\phi
_{i,0}e^{\delta }}\beta z\right) ^{2}}  \label{phii}
\end{equation}%
where $\phi _{i,0}=\phi _{i}(z=0)$ is the concentration at the
bottom. This expression is interesting because it shows
nonbarometric dependence as expected for the case of
nonaggregating objects.\cite{me}

The average aggregation number, $\overline{N}_{i}(z)=\sqrt{\phi
_{i}(z)e^{\delta }}$, has the hyperbolic dependence on the height $z$.\cite%
{Duyndam} Denoting the average aggregation number at the bottom, $\overline{N%
}_{i,0}=\overline{N}_{i}(z=0)$, it yields

\begin{equation}
\overline{N}_{i}(z)=\frac{\overline{N}_{i,0}}{1+\overline{N}_{i,0}\beta
z} \label{Ni}
\end{equation}

This expression shows, that the aggregates at the bottom are
longer than that on the top. This is simply a consequence of the
redistribution of subunits with height. It does not tell about the
change of the equilibrium size of the aggregates. In order to
examine the effect of gravity on the system as a whole, one can
consider the global average on the whole range of heights,
$\langle N_{i}\rangle $, as a function of $\beta =(mgh)/kT$.
\begin{figure}
\begin{center}
\includegraphics[height=5.9cm,width=8cm]{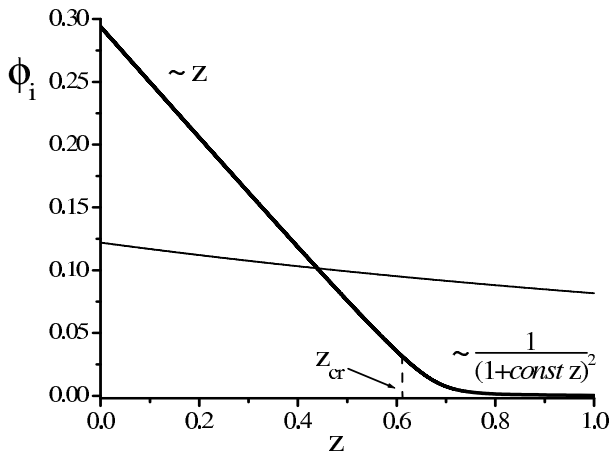}
\end{center}
\caption{Typical concentration profile of living polymers subject
to the gravitational field in a dilute solution (thick line) for
$\beta =0.7$, $ \delta =11$ and the total concentration $\Phi
_{i}=0.1$. The dependence $ \phi_{i}(z)$ is \textit{linear} except
the region on the top where the concentration is small. The
crossover between the regimes is $z_{cr}=\sqrt{ \Phi _{i}\pi
/\beta }$. In contrast, the concentration profile of
nonaggregating subunits ($\delta =0$) for the same total
concentration and the field (thin line) obeys the barometric
distribution.}
\end{figure}
This average is related to the total subunit concentration, $\Phi
_{i}=\int \phi _{i}(z)dz$, which assumes to be a constant

\begin{equation}
\langle N_{i}\rangle =\frac{\int NC_{i}(N,z)dNdz}{\int
C_{i}(N,z)dNdz}=\frac{ \Phi _{i}e^{\delta }}{\int
\overline{N}_{i}(z)dz}  \label{Ns}
\end{equation}

If the influence of the gravitational field is rather weak,
\textit{i.e.} when $\beta \ll \beta ^{\ast }\sim
1/\overline{N}_{i,0}$, it coincides with the expression in the
absence of the field, $\langle N_{i}\rangle \sim \sqrt{ \Phi
_{i}e^{\delta }}$. Thus, low gravitational field induces only the
redistribution of aggregates with height and it does not influence
the micellar equilibrium properties. On the other hand, if $\beta
\gtrsim \beta ^{\ast }$, $\langle N_{i}\rangle $
yields\cite{foot2}

\begin{equation}
\langle N_{i}\rangle =\frac{\Phi _{i}e^{\delta }\beta }{\ln \left[
1+\frac{1 }{2}\Phi _{i}e^{\delta }\beta ^{2}\left(
1+\sqrt{1+\frac{4}{\Phi _{i}e^{\delta }\beta ^{2}}}\right) \right]
}  \label{Niid}
\end{equation}
This is an increasing function of $\beta$, \textit{i.e.} strong
enough gravitational field is able to induce considerable
aggregation even of ideal micelles. The typical plot of this
function is presented in Fig. 1. Note, that this expression is
valid only for extremely low concentrations, $\Phi _{i}\ll 1$.
However, the deviation of this expression from the exact solution
(with the interaction term) is rather small. We can also calculate
the asymptote for $\beta \ll 1$,

\begin{equation}
\langle N_{i}\rangle \approx \sqrt{\Phi _{i}e^{\delta
}}+\frac{1}{24}\left( \Phi_{i}e^{\delta }\right)^{3/2}\beta
^{2}+O(\beta ^{4})
\end{equation}

\begin{figure}
\begin{center}
\includegraphics[height=5.9cm,width=8cm]{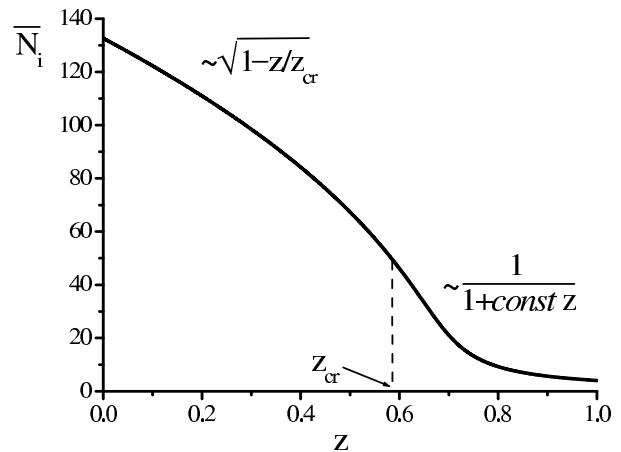}
\end{center}
\caption{The average aggregation number of living polymers in a
dilute solution, $\overline{N}_{i}(z)$, as a function of the
height $z$ for the same parameters as in Fig. 2. The plot can be
approximated by two asymptotes: long aggregates, interacting via
the second virial approximation for $z<z_{cr}$, and ideal
aggregates for $z>z_{cr}$ with $z_{cr}=\sqrt{\Phi _{i}\pi /\beta
}$.}
\end{figure}

If the concentration is not low, we can still get the analytical
result in the limit of \textit{long aggregates}, \textit{i.e.}
when we can neglect the third term in eq. (\ref{mui}). This can be
done for high enough end energies $\delta $, \textit{i.e.} in the
formal limit, $\delta \gg 1$. In this case, the concentration
depends linearly on the height,

\begin{equation}
\phi _{i}(z)=\phi _{i,0}-\frac{2}{\pi }\beta z  \label{phinoni}
\end{equation}
while the average aggregation number yields in the form

\begin{equation}
\overline{N}_{i}(z)=\sqrt{\overline{N}_{i,0}^{2}-\frac{2}{\pi
}e^{\delta }\beta z}  \label{Nilong}
\end{equation}%
where $\overline{N}_{i,0}=\sqrt{e^{\delta }(\Phi _{i}+\beta /\pi
)}$ is its value at the bottom, $z=0$. This expression brings us
immediately to the conclusion, that the interactions between
aggregates change the functional form of $\phi _{i}(z)$ and
$\overline{N}_{i}(z)$. In the limit of long
aggregates $\phi _{i}(z)$ is a linearly decreasing function and $\overline{N}%
_{i}(z)$ varies as a square root of height. Although the average
length of the aggregates can be hardly measured, the concentration
profile is easily accessible experimentally and the dependence of
$\phi _{i}(z)$ can be tested directly. The integration of eq.
(\ref{Ns}) gives the expression for the global average $\langle
N_{i}\rangle $ in the range of all heights in the form

\begin{equation}
\langle N_{i}\rangle =\frac{\Phi _{i}e^{\delta }\beta }{\frac{\pi
}{3} e^{\delta /2}\left[ \left( \Phi _{i}+\frac{\beta }{\pi
}\right) ^{3/2}-\left( \Phi _{i}-\frac{\beta }{\pi }\right)
^{3/2}\right] }
\end{equation}

Similarly to the case of ideal aggregates, $\langle N_{i}\rangle \sim \sqrt{%
\Phi _{i}e^{\delta }}$ for $\beta \ll \beta ^{\ast }\sim 1/\overline{N}%
_{i,0} $, while for strong fields it grows up with $\beta $,
indicating the increasing number of long aggregates:

\begin{figure}
\begin{center}
\includegraphics[height=6cm,width=8cm]{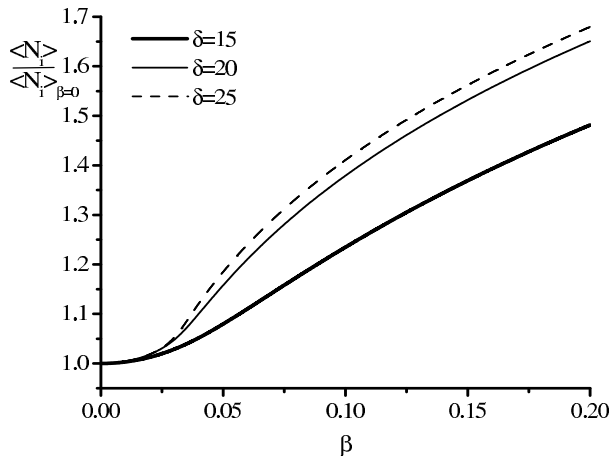}
\end{center}
\caption{Numerical plot of the global average aggregation number
on the whole range of heights, $\langle N_{i}\rangle $ of
\textit{nonideal} linear aggregates, rescaled to its value in the
absence of the field, $\langle N_{i}\rangle _{\beta =0}$, as a
function of $\beta =mgh/kT$ for different end energies: $\delta
=15$ (thick line, $\overline{N}_{i,0}\approx 200$), $\delta =25$
(thin line, $\overline{N}_{i,0}\approx 2000$), $\delta =25$ (dash
line, $\overline{N}_{i,0}\approx 25000$). The curve was calculated
by the integration of eq. (\ref{Ns}) for $\Phi _{i}=0.01$ (high
concentration).}
\end{figure}

\begin{equation}
\langle N_{i}\rangle =\sqrt{\Phi _{i}e^{\delta
}}+\frac{1}{24}\frac{\sqrt{\Phi _{i}e^{\delta }}}{\Phi _{i}\pi
^{2}}\beta ^{2}+O(\beta ^{4})
\end{equation}

It is noteworthy, that the increasing $\beta $ induces not only
the growth of the aggregates, it also leads to the precipitation
of the aggregates to the bottom. At some point, the subunit
concentration on the top will be close to zero and the
corresponding average aggregation numbers will not be high enough
to hold this approximation. To this end, the resulting curve of
the concentration profile, $\phi _{i}(z)$ and the average $\overline{N}%
_{i}(z)$ could be viewed as a combination of the above two limits. Namely, $%
\overline{ N}_{i}(z)\sim \sqrt{1-z/z_{cr}}$ at the bottom
($0<z<z_{cr}$) and $\overline{ N}_{i}(z)\sim (1+const$ $z)^{-1}$
on the top ($z_{cr}<z<1$). The crossover between the regimes,
$z_{cr}$, can be roughly considered as a
brake up of the approximation of long aggregates (\ref{Nilong}), $\overline{N%
}_{i}(z=z_{cr})=0$. That gives\cite{foot3}, $z_{cr}=\sqrt{\Phi
_{i}\pi /\beta }$. The concentration profile is presented in Fig.
2 and the average length corresponding to that profile depicted in
Fig. 3. Due to self-assembly, the influence of the gravitational
field on the concentration profile of living polymers is much
stronger than that for nonaggregating particles. This is an
important conclusion since concentration profiles can be measured
directly.

\begin{figure}
\begin{center}
\includegraphics[height=6.2cm,width=8cm]{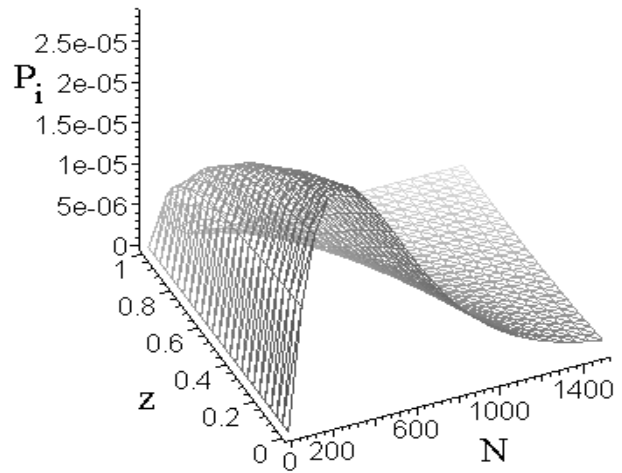}
\end{center}
\caption{Three dimensional plot of the distribution function of
the aggregate lengths at different heights in the isotropic phase
$P_{i}(N,z)$ (\ref{Pi}) for $\delta =15$, $\beta =0.03$, $\Phi
_{i}=0.01$. All the cross-sections with constant heights have the
same functional form.}
\end{figure}

The plot of $\langle N_{i}\rangle $ for the whole range of the
fields $\beta $ can be obtained numerically only. Eqs. $\mu
_{i}(z)=const$ (\ref{mui}) and the normalization condition $\Phi
_{i}=\int_{0}^{1}\phi _{i}(z)dz$ gives the concentration profile
as a function of the total subunit concentration $\Phi
_{i}$. The profile can be converted into the average length as $\overline{N}%
_{i}(z)=\sqrt{\phi _{i}(z)e^{\delta }}$ and inserted into eq.
(\ref{Ns}). The result is presented in Fig. 4. We plot the ratio
$\langle N_{i}\rangle /\langle N_{i}\rangle _{\beta =0}$, where
$\langle N_{i}\rangle _{\beta =0}$ is its value in the zero field.
This allows us to compare the relative growth of the aggregates
with different end energies, $\delta $. As we see, the aggregates
growth significantly in the gravitational field, especially for
large $\delta $.

At this point it is important to address the question of the
magnitude of the gravitational effect for a real system. The
typical example of microtubules used in the Introduction gives the
value of $\beta $ of the order of $0.001$, which is significantly
lower than the values in Fig. 4. These values can correspond to
rotating clinostat experiments\cite{clinostat} performed on
microtubules samples, where the system rotates about the
horizontal axis. The gravitational field in these experiments can be up to $%
300$ g. Moreover, the influence of the gravity on the global average $%
\langle N_{i}\rangle $ is noticeable even for normal gravitational
conditions in systems with low subunit concentrations: taking
$\Phi _{i}=10^{-5}$ and $\delta =30$ we get the average in the
absence of the gravity $\langle N_{i}\rangle _{\beta =0}\ $of the
order of $10000$, while for $\beta =0.001$ it yields $\langle
N_{i}\rangle \sim 16000$, \textit{i.e.} the normal gravitational
field induces considerable ($60\%$) growth of the aggregates.

However, $\langle N_{i}\rangle $ cannot show the redistribution of
aggregates of different lengths with height. In order to
illustrate the redistribution of aggregates in the gravitational
field, it is useful to consider the length distribution function
$P_{i}(N,z)=\int NC_{i}(N,z)d\Omega =4\pi NC_{i}(N,z)$. The eq.
(\ref{C}) gives

\begin{equation}
P_{i}(N,z)=N\exp \left( -\frac{N}{\overline{N}_{i}(z)}-\delta
\right) \label{Pi}
\end{equation}
where we assume $\overline{N}_{i}(z)$ in the form (\ref{Nilong}).
The three dimensional plot of this function is given in Fig. 5. It
is noteworthy, that
the envelope curve has the square root dependence in accordance with (\ref%
{Nilong}). The total number of aggregates, \textit{i.e.} the integral of $%
P_{i}(N,z)$ on $N$, is much higher at the bottom than on the top.
The maximum of the curve shifts to higher values of $N$ at the
bottom of the vessel.

\section{Nematic Ordering and Phase Equilibrium in the Gravitational Field}

Once the subunit concentration exceeds some critical value, the
nematic phase appears at the bottom of the vessel and the system
splits into two coexisting phases. It is known, that the average
length of the aggregates in
the absence of gravity is different in the nematic and the isotropic phases.%
\cite{McMullen,schootmain} In this chapter we focus on the
following questions: (i) how the gravitational field influence the
average length of both phases? (ii) is the growth in the nematic
phase as strong as in the isotropic phase? (iii) what would be the
distribution function on lengths of the aggregates in the
isotropic and the nematic phases as well as the change of the
overall distribution function of both phases with increasing
gravity? (iv) In particular, we will try to answer the question:
what is more favorable for an aggregate: to grow up in the
isotropic phase until some critical length and then to change to
the nematic phase, or to drop out into nematic phase immediately
in order to decrease its potential energy in the field? In any
case, it is quite obvious, that the aggregates in the isotropic
phase will shorten, while the aggregates in the nematic phase will
grow up with increasing gravity.

The simplest way to treat the nematic phase is to employ the
variational method\cite{Onsager,Odijk} that assumes the trial
orientational distribution function with the adjustable
variational parameter $\alpha$, which is found from the
minimization of the free energy. In analogy to the case in the
absence of gravity,\cite{schootisotropic,schootnem,Odijk} we may
write the interaction function $\rho _{n}(z)=4/\sqrt{\pi \alpha
(z)}$ and the entropy function $\sigma _{n}(z)=\ln (\alpha
(z)/4)+N\alpha (z)/(4p)$, where $p$ is the persistence length of
the rod per the length of one subunit.\cite{Odijk} The last term
reflects the flexibility of the rods. The only difference with
the expression in the absence of the field\cite{schootisotropic} is that $%
\alpha $ now depends on the height. Insertion of the expressions
for $\sigma _{n}(z)$ and $\rho _{n}(z)$ into (\ref{F}) and the
minimization with respect to $P(N,z)$ along with the constraint
$\int P_{n}(N,z)dN=\phi _{n}(z)$ gives the expression for the
distribution function $P_{n}(N,z)$ in the form

\begin{eqnarray}
&&\frac{P_{n}(N,z)}{N}=\exp\left(N\left[ \mu -\varepsilon _{\infty
}-\beta z- \right. \right. \nonumber \\
&& \left. \left. \frac{\pi }{2}\int P_{n}(N^{\prime },z)\rho
_{n}(z)dN^{\prime }\right] -\sigma _{n}(z)-\delta \right)
\end{eqnarray}
or applying the constraint (\ref{constr}) we can rewrite

\begin{equation}
\frac{P_{n}(N,z)}{N}=\frac{\phi _{n}}{\overline{N}_{n}^{2}(z)}\exp \left( -%
\frac{N}{\overline{N}_{n}(z)}\right)
\end{equation}
where the average aggregation number in the nematic phase yields

\begin{equation}
\overline{N}_{n}(z)=\sqrt{\frac{\alpha (z)}{4}\phi
_{n}(z)e^{\delta }} \label{Nsnz}
\end{equation}

\begin{figure}
\begin{center}
\includegraphics[height=6cm,width=8cm]{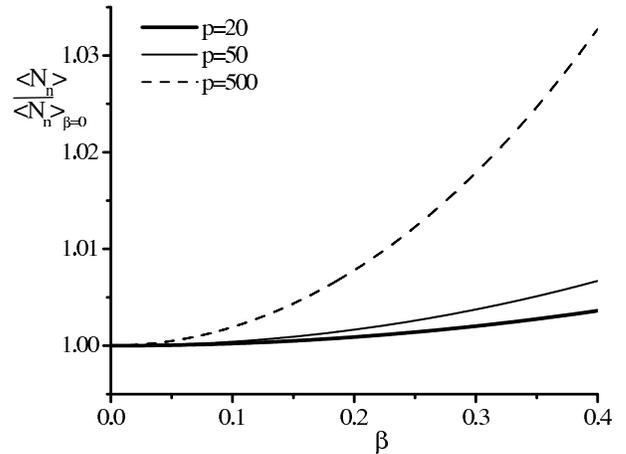}
\end{center}
\caption{The global average aggregation number on the whole range
of heights in the nematic phase, $\langle N_{n}\rangle $, eq.
(\ref{Nnn}), rescaled to its value in the absence of the field,
$\langle N_{n}\rangle _{\beta =0}$, as a function of $\beta
=mgh/kT$ with different flexibility of the aggregates: $p=20$
(thick line, $\overline{N}_{n,0}\approx 800$), $p=50$ (thin line,
$\overline{N}_{n,0}\approx 600$), $p=500$ (dash line,
$\overline{N}_{n,0}\approx 300$), for $\delta =15$, $\Phi
_{n}=0.6$. The overall growth is negligible with respect to the
isotropic phase, Fig. 4.}
\end{figure}

With this, the free energy of the nematic phase is

\begin{eqnarray}
\frac{F_{n}(z)}{kT}&=&\phi _{n}(z)\left[ \varepsilon _{\infty
}+\beta z+\frac{ \alpha (z)}{4p}-\frac{2}{\overline{N}_{n}(z)}+
\right. \nonumber \\
&&\left. \frac{\pi }{4}\phi _{n}(z) \frac{4}{\sqrt{\pi \alpha
(z)}}\right] \label{fn}
\end{eqnarray}

Minimization of the free energy with respect to $\alpha (z)$ in
the limit of considerably long rods, $\overline{N}_{n}(z)\gg 1$,
allows to relate $\alpha (z)$ and the local concentration as
\begin{equation}
\alpha (z)=\left( 2\sqrt{\pi }p\phi _{n}(z)\right) ^{2/3}
\label{alpha}
\end{equation}

The chemical potential $\mu _{n}(z)=\partial F_{n}(z)/\partial
\phi _{n}(z)$ is

\begin{equation}
\frac{\mu _{n}(z)}{kT}=\varepsilon _{\infty }+\beta z+\frac{\alpha
(z)}{4p}- \frac{1}{\overline{N}_{n}(z)}+\frac{\pi }{2}\phi
_{n}(z)\rho _{n}(z) \label{mun}
\end{equation}
and the osmotic pressure $p_{n}(z)=\phi _{n}^{2}/\upsilon
\partial \left[ F_{n}/\phi _{n}\right] /\partial \phi _{n}(z)$ is

\begin{equation}
\frac{p_{n}(z)}{kT}\upsilon =\phi _{n}(z)\left[
\frac{1}{\overline{N}_{n}(z)} +\frac{\pi }{4}\phi _{n}(z)\rho
_{n}(z)\right]  \label{pn}
\end{equation}
The form of these expressions is close to that in the absence of gravity.%
\cite{schootisotropic}

\begin{figure}
\begin{center}
\includegraphics[height=6cm,width=8cm]{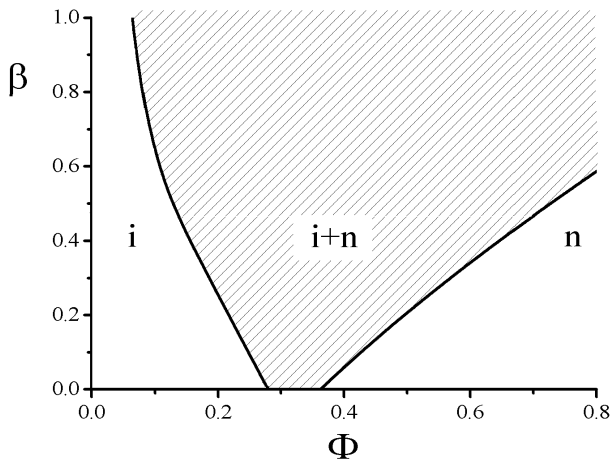}
\end{center}
\caption{Phase diagram of living polymers in the gravitational
field in the variables: the total subunit concentration $\Phi $
and the parameter associated with the gravitational field $\beta
=mgh/kT$ for the parameters $p=40$, $\delta =10$ leading to the
phase transition even in the absence of gravity. The
concentrations corresponding to the onset of the phase separation
diverge and the coexistence region becomes broader. Here i denotes
the isotropic phase, n is the nematic phase, i+n corresponds to
the phase coexistence region.}
\end{figure}

Let us first calculate the average aggregation number,
$\overline{N}_{n}(z)$ , in the limit of long aggregates,
\textit{i.e.} when $\delta $ is large, so we can neglect the
fourth term in (\ref{mun}). Noting that $\frac{\pi }{2} \phi
_{n}(z)\rho _{n}(z)=\alpha (z)/p$, the chemical potential
(\ref{mun}) is

\begin{equation}
\frac{\mu _{n}(z)}{kT}\approx \varepsilon _{\infty }+\beta
z+\frac{5}{4} \frac{\alpha (z)}{p}
\end{equation}

Thus, $\alpha (z)$ changes linearly with height, $\alpha
(z)=\alpha _{n,0}-4/5p\beta z$, where $\alpha _{n,0}=\alpha
(z=0)$. The local concentration, is then $\phi _{n}(z)=\left( \phi
_{n,0}^{2/3}-4/5p^{1/3}\beta z/(2\sqrt{\pi })^{2/3}\right) ^{3/2}$ with $%
\phi _{n,0}$ being the concentration at the bottom, and
correspondingly, the average aggregation number is

\begin{equation}
\overline{N}_{n}(z)=\left[ \overline{N}_{n,0}^{4/5}-\frac{4}{5}\left( \frac{%
e^{\delta }}{8\sqrt{\pi }p}\right) ^{2/5}p\beta z\right] ^{5/4}
\label{Nn}
\end{equation}
where $\overline{N}_{n,0}\equiv \overline{N}_{n}(z=0)=\frac{1}{2}\sqrt{%
\alpha _{n,0}\phi _{n,0}e^{\delta }}$. Hence, we can conclude that $%
\overline{N}_{n}(z)$ changes almost linearly with height in
contrast to the square root behavior in the isotropic phase. Since
the nematic phase is composed mostly of long aggregates, the
approximation of long aggregates works well in a large range of
concentrations.

\begin{figure}
\begin{center}
\includegraphics[height=6cm,width=8cm]{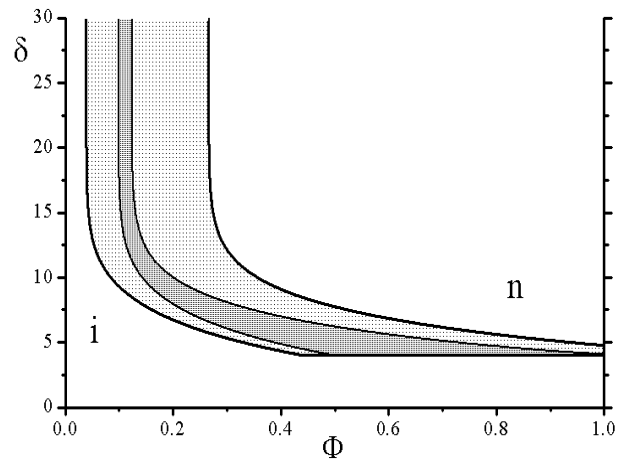}
\end{center}
\caption{Phase diagram of living polymers in the gravitational
field in the variables: the total subunit concentration $\Phi $
and the end energy of linear aggregates $\delta $ for $p=100$. The
dark pattern corresponds to the phase separation region in the
absence of the gravity, $\beta =0$, while the light pattern
corresponds to the gravitational field $\beta =0.2$, i denotes the
isotropic phase, n denotes the nematic phase. The phase separation
region becomes broader in the gravitational field.}
\end{figure}

Assuming, that the vessel is entirely occupied by the nematic
phase, we can calculate the global average length of the
aggregates on the whole range of heights

\begin{equation}
\langle N_{n}\rangle =\frac{\Phi _{n}e^{\delta }}{\int
\overline{N}_{n}(z)dz} \label{Nnn}
\end{equation}

Although an analytical expression for this case is not possible,
we can plot the numerical result. Integrating eq. (\ref{Nn}) and
fixing the total concentration $\Phi _{n}=\int_{0}^{1}\phi
_{n}(z)dz$, we get with the help of eqs. (\ref{Nsnz}) and
(\ref{alpha}) the dependence of $\langle N_{n}\rangle $ on $\beta
$ (Fig. 6). Here again we plot it as a rescaled variable, $\langle
N_{n}\rangle /\langle N_{n}\rangle _{\beta =0}$, in order to
compare the growth of aggregates with different stiffness, $p$. As
in the case of the isotropic phase, the gravitational field leads
to the growth of the aggregates, though the growth in the nematic
phase is less pronounced. One can see that stiff aggregates ($p$
is large) growth better than flexible ones ($p$ is small). By
further increasing $\beta $ the top of the vessel rarefies until
the appearance of the isotropic phase coexisting with the nematic
phase at the bottom.

To treat the phase separation region we can employ the same method
as in ref. \onlinecite{me} which is valid for the coexisting
phases with concentration gradients. Once a system is subject to a
concentration gradient, there is no concentration which determines
the equilibrium properties of a phase. Instead, we will consider
the \textit{average} concentration in the nematic phase, $\Phi
_{n}$, which is assumed to be at the bottom, $z<x$, and the
\textit{average} concentration in the isotropic phase, $\Phi
_{i}$, at the top $z>x$. where $x$ is the height of the phase
boundary:

\begin{figure}
\begin{center}
\includegraphics[height=6cm,width=8cm]{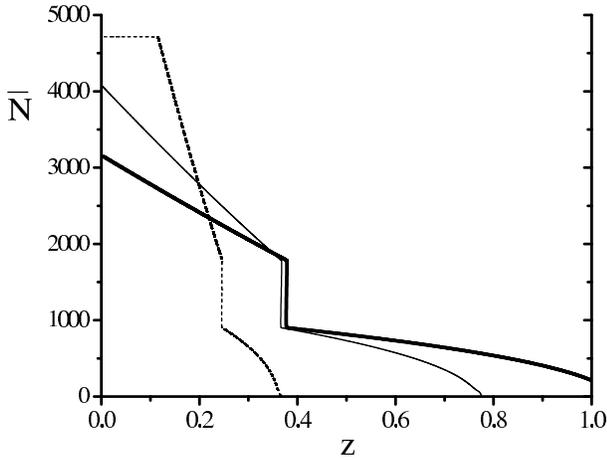}
\end{center}
\caption{Average aggregation number, $\overline{N}(z)$, as a
function of height, $z$, in the phase coexistence region for three
different parameters associated with the gravitational field,
$\beta =0.6$ (thick solid curve), $\beta =1.0$ (thin solid curve),
$\beta =3.5$ (dash curve). The jump corresponds to the position of
the phase boundary, $x$, the nematic phase is at the bottom,
$0<z<x$, and the isotropic phase is on the top, $x<z<1$. The
parameters used: $p=40$, $\delta =15$, $\Phi =0.255$.}
\end{figure}

\begin{eqnarray}
\Phi _{n} &=&\frac{1}{x}\int_{0}^{x}\phi _{n}(z)dz  \label{Phin} \\
\Phi _{i} &=&\frac{1}{1-x}\int_{x}^{1}\phi _{i}(z)dz  \label{Phii}
\end{eqnarray}
The concentrations are related via the conservation of mass
condition $\Phi =x\Phi _{n}+(1-x)\Phi _{i}$, where $\Phi $ is the
total subunit concentration in both phases. Note, that $\Phi _{n}$
and $\Phi _{i}$ are not fixed, they are determined from the
equilibrium conditions.

The equilibrium corresponds to the equality of osmotic pressures
and chemical potentials at the boundary, $z=x$. Using the
expressions for the
isotropic phase (\ref{mui}), (\ref{pi}) and the nematic phase, (\ref{mun}), (%
\ref{pn}), we can write

\begin{eqnarray}
-\frac{1}{\sqrt{\phi _{i}(x)e^{\delta }}}+\frac{\pi }{2}\phi
_{i}(x) &=& \frac{\alpha (x)}{4p}-\frac{1}{\sqrt{\frac{1}{4}\alpha
(x)\phi _{n}(x)e^{\delta }}}+ \nonumber \\
&&\frac{\pi }{2}\phi _{n}(x)\frac{4}{\sqrt{\pi \alpha (x)}
}  \label{eq1} \\
\frac{\phi _{i}(x)}{\sqrt{\phi _{i}(x)e^{\delta }}}+\frac{\pi
}{4}\phi _{i}^{2}(x) &=&\frac{\phi
_{n}(x)}{\sqrt{\frac{1}{4}\alpha (x)\phi _{n}(x)e^{\delta }}}+
\nonumber \\
&&\frac{\pi }{4}\phi _{n}^{2}(x)\frac{4}{\sqrt{\pi \alpha (x)}}
\label{eq2}
\end{eqnarray}
where $\alpha (z)$ is determined from eq. (\ref{alpha}). The
solution of four equations (\ref{Phin})-(\ref{eq2}) along with the
conservation of mass condition, $\Phi =x\Phi _{n}+(1-x)\Phi _{i}$,
gives the equilibrium distribution of aggregates between two
phases.
\begin{figure}
\begin{center}
\includegraphics[height=6cm,width=8cm]{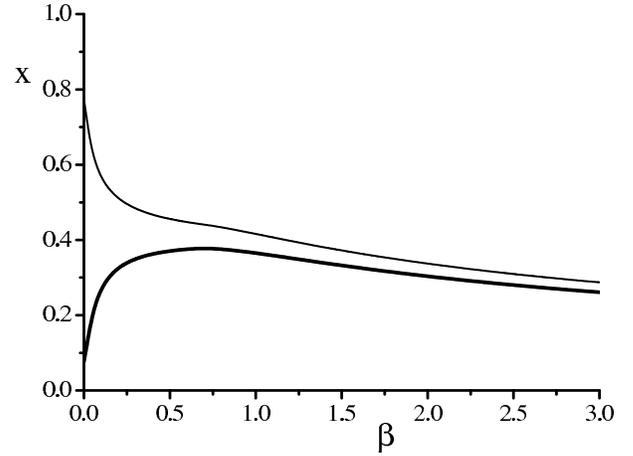}
\end{center}
\caption{The position of the boundary between two phases, $x$, as
a function of $\beta =mgh/kT$ for two different total subunit
concentrations: $\Phi =0.255$ (thick line) and $\Phi =0.3$ (thin
line). The parameters used: $\delta=15$, $p=40$.}
\end{figure}
Note, that the last two
equations determine the boundary concentrations of the coexisting phases, $%
\phi _{i}(x)$ and $\phi _{n}(x)$. The form of the equations
coincides with that in the absence of
gravity,\cite{schootisotropic} since the aggregates at the
boundary have all the same potential energy and the gravitational
field does not influence their behavior. Chemical potentials in
both phases do not depend on the height. They can be regarded as
equations for the definition of local concentrations $\phi
_{i,n}(z)$ as implicit functions of height. To get the average
concentrations of coexisting phases, $\Phi _{n}$ and $\Phi _{i}$,
as well as the boundary position, $x$, we have to integrate the
local concentrations. However, it only leads to implicit
definition of local concentrations. This results in a system of
integral equations which can be hardly solved. See Appendix how to
avoid this difficulty.

In contrast to the phase separation in a homogeneous system, the
coexisting phases in a gravitational field exhibit the
concentration gradient and so, the average concentrations of the
phases,$\Phi _{i}$ and $\Phi _{n}$, \textit{do not correspond} to
the onset concentrations of the phase
transition. This is because they both depend on the total concentration, $%
\Phi$. In order to find the values of the total concentration
signifying the phase separation boundaries, we put in the
equations $x=0$ for the onset of the isotropic phase and $x=1$ for
the onset of the nematic phase. Solving the above equations with
respect to $\Phi $, the boundaries of the coexistence region can
be obtained. Fig. 7 represents the typical phase diagram in the
variables $\beta -\Phi $. The two curves diverges, thus, the
gravitational field significantly broaden the coexistence region.
Starting at some point in the isotropic phase at zero field and
moving vertically on the plot, \textit{i.e.} increasing the field,
at some $\beta $ we enter the phase separation region: increasing
local concentration at the bottom reaches the limiting value and
the nematic phase appear at the bottom. Increasing gravity in the
nematic phase leads to the rarefaction on the top, what gives rise
to the isotropic phase. Thus, the gravity facilitates the
isotropic to nematic phase transition.

\begin{figure}
\begin{center}
\includegraphics[height=6cm,width=8cm]{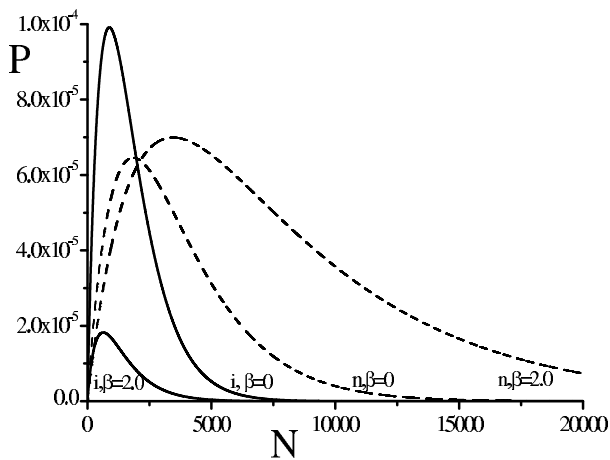}
\end{center}
\caption{Length distribution functions in the nematic phase,
$P_{n}(N)$ (dash curves), and in the isotropic phase, $P_{i}(N)$
(solid curves), for two cases: in the absence of the field,
subscript $\beta =0$, and in the field, subscript $\beta =2.0$.
The parameters used: $p=40$, $\delta =15$, $\Phi =0.255$.}
\end{figure}

Since we describe the self-assemble system, which is sensitive to
the variational of external conditions, in particular to the
change of the temperature, it is useful to plot the phase diagram
in the variables $\delta -\Phi $, where the end energy, $\delta $,
usually has the inverse temperature dependence. The result is
presented in Fig. 8. Here the inner phase diagram (dark pattern)
is the coexistence region in the absence of the field in
accordance with refs.\cite{taylor,schootnem} As we see, to induce
the phase separation, $\delta $ should exceed some critical value,
for small $\delta $ the aggregates are too short and the nematic
phase never appears.
(Compare it with theoretically predicted and experimental phase diagrams.)%
\cite{taylor,Herzfeld} The phase diagram depends strongly on the
stiffness parameter $p$. In the limit of rigid rods, $p\rightarrow
\infty $, the concentrations of coexisting phases tend to Y-axis,
while for flexible aggregates, there is a little gap,
\textit{i.e.} stiff aggregates much easily form the nematic phase
than the flexible ones. The gravitational field severely increases
the coexistence region (light pattern) in both cases. So, for
strong enough gravitational fields the left border can approach
the Y-axis, signifying the absence of the pure isotropic phase for
high values of $\delta $ even for very low concentrations. This
situation should be taken into account, when the self-assembled
aggregates are subject to the ultracentrifugal field.

\begin{figure}
\begin{center}
\includegraphics[height=6cm,width=8cm]{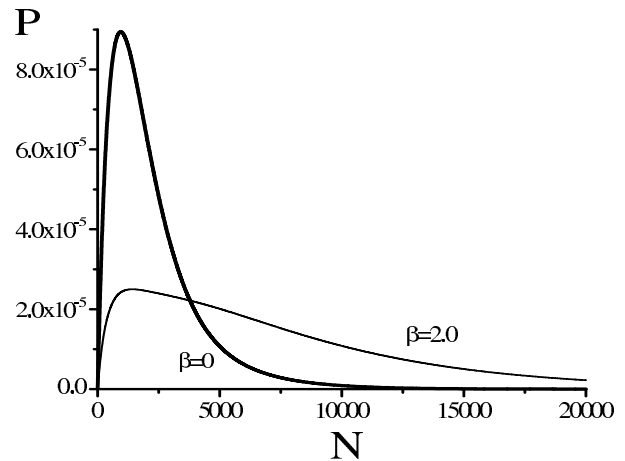}
\end{center}
\caption{The sum of the length distribution functions of the
nematic and isotropic phases, presented in Fig. 11,
$P(N)=xP_{n}(N)+(1-x)P_{i}(N)$ for $\beta =0$ (thick line) and
$\beta =2.0$ (thin line). The parameters used: $p=40$, $\delta
=15$, $\Phi =0.255$.}
\end{figure}

Let us focus on the coexistence region in more details. The
solution of four equations (\ref{Phin})-(\ref{eq2}) along with the
conservation of mass condition for each $\beta $ gives not only
the average concentrations in the coexisting phases, it also gives
the position of the phase boundary, $x$, and the concentrations at
the boundary, $\phi _{i}(x)$ and $\phi _{n}(x)$. This information
is enough to obtain the full concentration profile composed of two
parts, $\phi _{n}(z)$ at $0<z<x$ and $\phi _{i}(z)$ at $x<z<1$.
This profiles can be converted into the average aggregation
numbers in both phases:
\begin{eqnarray}
\overline{N}_{i}(\phi _{i}(z)) &=&\sqrt{\phi _{i}(z)e^{\delta }}
\label{Nnphii} \\
\overline{N}_{n}(\phi _{n}(z)) &=&\sqrt{\left( 2\sqrt{\pi
}p\right) ^{2/3}\phi _{n}^{5/3}(z)e^{\delta }/4}  \label{Nnphinn}
\end{eqnarray}
The resulting plot for different $\beta $ is shown in Fig. 9
(Actually, we plot $z_{i,n}(\phi )$ and thus, $z_{i,n}(N)$ as
explained in Appendix). The average aggregation number in the
isotropic phase exhibits the square root behavior on height, and
it has almost linear dependence in the nematic phase in accordance
to our previous estimates. The only exception is the region at
the very bottom for high gravitational field $\beta =3.5$, where $\overline{N%
}_{n}$ does not depend on $z$. This is because the concentration
at the bottom for high gravitational fields is very high and the
second virial approximation does not work. Thus, after correcting
this approximation, the sharp edge on the dashed curve should
smooth out. The length of the concentration jump at the phase
boundary does not depend on the field, since the field can only
shift the phase boundary and it does not influence the equilibrium
of the aggregates lying in the same height.

As in the case of monodisperse rods of fixed length, the gravity
can either increase or decrease the volume of the nematic phase
depending on the total subunit concentration (Fig. 10). If the
total concentration is close to the isotropic phase boundary and
the volume of the nematic phase is low, the gravitational field
induces the increase of the nematic phase, so the volume increases
until exhausting of the isotropic phase. Since then, the volume
gradually decreases indicating the sinking of the aggregates in
the nematic phase (solid line). If the total concentration is
close to the nematic phase boundary, the aggregates in the nematic
phase can only sink, decreasing the volume (thin line).

Now, let us address the question of the \textit{polydispersity} of
the nematic and the isotropic phases in the gravitational field.
The length
distribution functions of aggregates in coexisting phases, $P_{i}(N)$ and $%
P_{n}(N)$ can be calculated as

\begin{eqnarray}
P_{i}(N) &=&\frac{N}{1-x}\int_{x}^{1}\frac{\phi
_{i}(z)}{\overline{N} _{i}^{2}(z)}\exp \left(
-\frac{N}{\overline{N}_{i}(z)}\right) dz  \label{Pi1}
\\
P_{n}(N) &=&\frac{N}{x}\int_{0}^{x}\frac{\phi
_{n}(z)}{\overline{N} _{n}^{2}(z)}\exp \left(
-\frac{N}{\overline{N}_{n}(z)}\right) dz  \label{Pn1}
\end{eqnarray}
The result is presented in Fig. 11. See Appendix and especially eqs. (\ref%
{APNi}), (\ref{APNn}) for the details of the calculation. The
gravity induces the transition of subunits from the isotropic to
the nematic phase. So, the area of the isotropic distribution
function, $P_{i}(N)$, decreases with increasing gravity, while the
area of $P_{n}(N)$ increases with $\beta $. Nevertheless, the
distribution of lengths in the nematic phase is much broader in
the presence of the field, \textit{i.e.} the gravity increases
seriously the polydispersity of the nematic phase. This is for
account of the fraction of extremely long aggregates induced by
the gravity, while the fraction of short aggregates remains almost
unchanged. The isotropic phase looses foremost long, and thus,
heavy aggregates, but the change in its polydispersity is not so
drastic as in the nematic phase. Hence, \textit{the precipitation
into nematic phase is more favorable process than the growth in
the isotropic phase}. To describe the polydispersity of the whole
two
phase system, we can consider the sum of the length distribution functions, $%
P=xP_{n}(N)+(1-x)P_{i}(N)$. The area of this function is
proportional to the
total number of subunits in the system, so its value is the same for any $%
\beta$. The plot of $P(N)$ is shown in the Fig. 12. It brings us
to the conclusion that the system as a whole is more polydisperse
in the gravitational field. Sharply picked function in the absence
of gravity fade out with increasing gravitational field. Note,
however, that the maximum of the curve almost does not change.
This is the consequence of the redistribution of subunits between
two phases without considerable growth of the aggregates.

\section{Conclusion}

Although the gravitational field is often neglected in the
description of physicochemical processes, it may have noticeable
influence on such biological heavy objects as microtubules or
actin filaments, essential components of living cells. The
equilibrium properties of such objects is influenced even by the
earth gravitational acceleration. The fields in an ultracentrifuge
can multiple the effect of gravity.

In this paper we focus on the major effects of the gravitational
field on the process of self-assembly of linear aggregates. It is
shown, that the concentration gradient induced by the
gravitational field leads to the redistribution of aggregates of
different lengths (and thus, masses) on different heights: the
local concentration of the self-assembled rods interacting via the
second virial coefficient decreases \textit{linearly} with height
in contrast to barometric distribution of aggregates with fixed
length. This prediction can be directly checked experimentally in
the rotating clinostat experiments. In turn, the average length of
the aggregates has a square root dependence in contrast to
hyperbolic dependence found for ideal aggregates.\cite{Duyndam} We
also show, that except the
redistribution of aggregates with height, the gravity induces the \textit{%
overall growth} of the aggregates. The stiffness of the aggregates
enhance the growth under gravity. Strong enough gravitational
field can provoke the isotropic to nematic phase transition, which
is reflected in the significant broadness of the phase coexistence
region. The transition manifests itself in the precipitation of
the aggregates from the isotropic to the nematic phase regardless
their length. Furthermore, the length distribution of the
aggregates is found to be very sensitive to the gravitational
field: the gravity induces considerable polydispersity in the
nematic phase. Taken together, the effect of gravity can lead to a
series of experimentally relevant observations which are
interesting from fundamental point of view.

\newpage \appendix*

\section{'}

Equations $\mu _{i,n}=const$ (\ref{mui}), (\ref{mun}) do not allow
to obtain the analytical expression of local concentrations $\phi
_{i,n}(z)$ as a function of $z$ except for two asymptotes eqs.
(\ref{phii}), (\ref{phinoni}). To this end, the total
concentration, $\Phi =\int \phi (z)dz$, and the average lengths,
$\overline{N}_{i,n}(z)$, are, in fact, the integral equations for
the functions $\phi _{i,n}(z)$, the functional forms of which are
determined from the chemical potentials. In general, it is rather
difficult to solve such system of equations. However, we can
employ the structure of equations (\ref{mui}), (\ref{mun}), which
allows to get analytically the \textit{inverse} functions,
$z_{i,n}(\phi )$. Namely, for the isotropic phase

\begin{equation}
z_{i}(\phi )=x-\frac{1}{\beta }\left[ \frac{1}{\sqrt{\phi
_{i}(x)e^{\delta }} }-\frac{1}{\sqrt{\phi e^{\delta }}}-\frac{\pi
}{2}\left( \phi _{i}(x)-\phi \right) \right]
\end{equation}

Here we assume the isotropic phase is situated above the phase
boundary, $x<z_{i}<1$, and the boundary concentration is $\phi
_{i}(x)$. In case of the homogeneous phase, one can put in this
expression $x=0$. Analogously, the inverse function for the
nematic phase yields in the form

\begin{eqnarray}
z_{n}(\phi ) &=&x+\frac{5\pi ^{1/2}e^{\delta /2}\left( \phi
_{n}^{2/3}(x)-\phi ^{2/3}\right) }{2^{4/3}\pi
^{1/6}p^{1/3}e^{\delta
/2}\beta }+  \nonumber \\
&&\frac{4\left( 1/\phi ^{5/6}-1/\phi _{n}^{5/6}(x)\right)
}{2^{4/3}\pi ^{1/6}p^{1/3}e^{\delta /2}\beta }
\end{eqnarray}

The two expressions should be accompanied by the boundary conditions, $%
z_{i}(\phi =\phi _{i}(1))=1$ and $z_{n}(\phi =\phi _{n}(0))=0$.

With this, the average concentrations in the phases are

\begin{eqnarray}
\Phi _{i} &=&\frac{1}{1-x}\int_{\phi _{i}(x)}^{\phi _{i}(1)}\phi
\frac{
\partial z_{i}(\phi )}{\partial \phi }d\phi  \label{APhii} \\
\Phi _{n} &=&\frac{1}{x}\int_{\phi _{n}(0)}^{\phi _{n}(x)}\phi
\frac{
\partial z_{n}(\phi )}{\partial \phi }d\phi  \label{APhin}
\end{eqnarray}
where the value of the local concentration is $\phi _{n}(0)$ at
the bottom and $\phi _{i}(1)$ on the top.

The derivatives in the integrals are

\begin{eqnarray}
\frac{\partial z_{i}(\phi )}{\partial \phi } &=&-\frac{\pi
}{2\beta }\left(
1+\frac{1}{\pi \phi \sqrt{\phi e^{\delta }}}\right) \\
\frac{\partial z_{n}(\phi )}{\partial \phi }
&=&-\frac{5}{3}\frac{1}{\left( 2 \sqrt{\pi }p\right) ^{1/3}\phi
^{11/6}\beta }\left( \sqrt{\pi }\phi ^{3/2}+e^{-\delta /2}\right)
\end{eqnarray}

Thus, the average concentrations can be integrated analytically.

The global length average of the aggregates in both phases,
(\ref{Ns}) and ( \ref{Nnn}), can be written as integrals on $\phi
$:

\begin{eqnarray}
\langle N_{i}\rangle &=&\Phi _{i}e^{\delta /2}\left/ \int_{\phi
_{i}(x)}^{\phi _{i}(1)}\phi ^{1/2}\frac{\partial z_{i}(\phi
)}{\partial \phi
}d\phi \right.  \label{APNi} \\
\langle N_{n}\rangle &=&\frac{\Phi _{n}e^{\delta /2}}{\left(
2\sqrt{\pi } p\right) ^{1/3}}\left/ \int_{\phi _{n}(0)}^{\phi
_{n}(x)}\phi ^{5/6}\frac{
\partial z_{n}(\phi )}{\partial \phi }d\phi \right.  \label{APNn}
\end{eqnarray}
The length distribution functions (\ref{Pi1}), (\ref{Pn1}) yield
in the form

\begin{eqnarray}
P_{i}(N) &=&\frac{N}{1-x}e^{-\delta }\int_{\phi _{i}(x)}^{\phi
_{i}(1)}\exp
\left( -\frac{N}{\sqrt{\phi e^{\delta }}}\right) \frac{\partial z_{i}(\phi )%
}{\partial \phi }d\phi  \\
P_{n}(N) &=&\frac{N}{x}\frac{4e^{-\delta }}{\left( 2\sqrt{\pi
}p\right)
^{2/3}}\int_{\phi _{n}(0)}^{\phi _{n}(x)}\phi ^{-2/3}\times   \nonumber \\
&&\exp \left( -\frac{2N}{\sqrt{\left( 2\sqrt{\pi }p\right)
^{2/3}\phi ^{5/3}e^{\delta }}}\right) \frac{\partial z_{n}(\phi
)}{\partial \phi }d\phi
\end{eqnarray}
where we used (\ref{Nnphii}), (\ref{Nnphinn}) for the averages in
the isotropic and the nematic phases.

Solution of \emph{algebraic} eqs. (\ref{eq1}), (\ref{eq2}), (\ref{APhii}), ( %
\ref{APhin}) along with the conservation mass condition, $\Phi
=x\Phi _{n}+(1-x)\Phi _{i}$, gives the phase diagram as well as
the equilibrium average concentrations in the coexisting phases.


\begin{thebibliography}{99}

\bibitem{catescando} Cates M. E., Candau S. J., J. Phys.: Condens. Matter,
\textbf{2}, 6869 (1990).

\bibitem{Greer} Greer S., in \emph{Advances in Chem. Phys.}, XCIV, Polymeric
Systems, edited by Prigogine I. and Rice S., (Interscience, N.Y.
1996).

\bibitem{Shadows} Penrose, R. \emph{Shadows of the Mind} (Oxford University
Press, N.Y. 1994).

\bibitem{DogteromLes} Dogterom M. in \emph{Physics of Bio-molecules and Cells%
}, edited by Flyvbjerg H, Julicher F., Ormos P., David F., Les
Houches Session LXXV (EDP Sciences, Springer-Verlag, 2002).

\bibitem{Dogterom} Dogterom M., Yurke B., Science, \textbf{278}, 856 (1997).

\bibitem{mitosis} Scholey J. M., Rogers G. C., Sharp D.J., The Journal of
Cell Biology, \textbf{154}(2), 261 (2001).

\bibitem{Panda} Panda D., Chakrabarti G., Hudson J., Pigg K., Miller H.,
Wilson L., Himes R., Biochemistry, \textbf{39}, 5075 (2000).

\bibitem{Derry} Derry W., Wilson L., Khan I., Luduena R., Jordan M. A.,
Biochemistry, \textbf{36}, 3554 (1997).

\bibitem{Anacker} Anacker E., Rush R., Johnson J., J. Phys. Chem., \textbf{%
68 }(1), 81 (1964).

\bibitem{Shah} Shah C., Zhi-Qi Xu C., Vickers J., Williams R., Biochemistry,
\textbf{40}, 4844 (2001).

\bibitem{Hlavanda} Hlavanda E., Kovacs J., Olah J., Orosz F., Medzihradszky
K., Ovadi J., Biochemistry, \textbf{41}, 8657 (2002).

\bibitem{clinostat} Tabony J., Glade N., Papaseit C., Demongeot J., Cell
Biology and Biotechnology in Space, 19 (2002) and Glade N., Tabony
J., J. Phys. IV France, \textbf{11}, Pr6-255 (2001).

\bibitem{Tabony} Tabony J., Glade N., Demongeot J., Papaseit C., Langmuir,
18, 7196 (2002).

\bibitem{tabony2} Tabony J., Glade N., Papaseit C., Demongeot J., J. Phys.
IV France \textbf{11}, Pr6-239 (2001).

\bibitem{tabony3} Tabony J., Pochon N., Papaseit C., Advances in Space
Research, \textbf{28}(4), 529 (2001).

\bibitem{liposomes} Claassen D., Spooner B., Advances in Space Research,
\textbf{17}(6-7), 151 (1995).

\bibitem{micrograv} Dag O., Ahari H., Coombs N., Jiang T., ArocaOuellette
P., Petrov S., Sokolov I., Verma A., Vovk G., Young D., Ozin G.,
Reber C., Pelletier Y., Bedard R., Advanced Materials,
\textbf{9}(15), 1133 (1997).

\bibitem{crystal} Hong J., J. Crystal Growth, \textbf{181}, 459 (1997) and
references therein.

\bibitem{Duyndam} Duyndam A., Odijk T., Langmuir, \textbf{9}, 1160 (1993).

\bibitem{Duyndam2} Duyndam A., Odijk T., Langmuir, \textbf{14}, 2577 (1998).

\bibitem{Fraden} Dogic Z., Philipse A., Fraden S., Dhont J., J. Chem. Phys.,
\textbf{113}(18), 8368 (2000).

\bibitem{me} Baulin V.A., Khokhlov A.R., Phys. Rev. E, \textbf{60}(3), 2973
(1999).

\bibitem{Safran} Safran, S. A. \emph{Statistical Thermodynamics of Surfaces,
Interfaces and Membranes} (Addison-Wesley, Reading, MA, 1994).

\bibitem{OdijkRev} Odijk T., Curr. Opin. in Col. \& Int. Sci., \textbf{1},
337 (1996).

\bibitem{taylor} Taylor M., Herzfeld J., J. Phys.: Condens. Matter, \textbf{5},
2651 (1993).

\bibitem{gelbart1} Gelbart W., Ben-Shaul A., McMullen W., Masters A., J.
Phys. Chem., \textbf{88}, 861 (1984).

\bibitem{gelbart2} Gelbart W., McMullen W., Ben-Shaul A., J. Physique,
\textbf{46}, 1137 (1985).

\bibitem{McMullen} McMullen W., Gelbart W., Ben-Shaul A., J. Chem. Phys.,
\textbf{82}(12), 5616 (1985).

\bibitem{Cuesta} Cuesta J., Sear R., Eur. Phys. J. B, \textbf{8}, 233 (1999).

\bibitem{schootmain} van der Schoot P., Cates M., Langmuir, \textbf{10}, 670 (1994).

\bibitem{schootgrowth} van der Schoot P., Europhys. Lett., \textbf{39}(1), 25 (1997).

\bibitem{schootphase} van der Schoot P., J. Phys. II (France), \textbf{5}, 243
(1995).

\bibitem{schootisotropic} van der Schoot P., Cates M., Europys. Lett., \textbf{25}
(7), 515 (1994).

\bibitem{schootnem} van der Schoot P., Macromolecules, \textbf{27}, 6473 (1994).

\bibitem{Odijk} Odijk T., Macromolecules, \textbf{19}, 2313 (1986).

\bibitem{Onsager} Onsager L., Ann. (N.Y.) Acad. Sci., \textbf{51}, 627
(1949).

\bibitem{foot1} In the isotropic phase the integral in the exponent of eq. (%
\ref{CNOm}) can be calculated directly. It yields in the form
$2/\upsilon \int C(N,\Omega ,z)B_{2}(\gamma )dNd\Omega =\pi /2\phi
N$. Here the equality $\int \left\vert \sin \gamma \right\vert
d\Omega =\pi ^{2}$ was used.

\bibitem{foot2} Integration of (\ref{phii}) on the whole range of $z$ allows
to relate $\overline{N}_{i,0}$ and $\Phi _{i}$. With this, the
substitution
of (\ref{Ni}) into eq. (\ref{Ns}) and subsequent integration gives (\ref%
{Niid}).

\bibitem{foot3} To calculate $z_{cr}$ we assume that the subunits
concentration for $z>z_{cr}$ is negligibly small and $\Phi
_{i}=\int_{0}^{z_{cr}}\phi _{i}(z)dz$. Integration and solution of
this
equation with respect to $\overline{N}_{i,0}$ gives $\overline{N}_{i,0}=%
\sqrt{e^{\delta }\left( \Phi _{i}+z_{cr}^{2}\beta /\pi \right)
/z_{cr}}$. Substitution of this result into (\ref{Nilong}) and
solution of the
subsequent equation $\overline{N}_{i}(z)=0$ gives the crossover height $%
z_{cr}=\sqrt{\Phi _{i}\pi /\beta }$.

\bibitem{Herzfeld} Herzfeld J., Acc. Chem. Res., \textbf{29}, 31 (1996).


\end{thebibliography}
\end{document}